\documentclass[10pt,twocolumn]{article} 
\usepackage{simpleConference}
\usepackage{times}
\usepackage{graphicx}
\usepackage{gensymb}
\usepackage{amssymb}
\usepackage{amsmath}
\usepackage{url,hyperref}
\usepackage{subfigure}
\usepackage{bm}
\usepackage{upgreek}

\begin{document}

\title{Continuous head-related transfer function representation based on hyperspherical harmonics}

\author{Adam Szwajcowski \\
\\
AGH University of Science and Technology \\ Department of Robotics and Mechatronics
\\
szwajcowski@agh.edu.pl  \\
}

\maketitle
\thispagestyle{empty}

\begin{abstract}
Expressing head-related transfer functions (HRTFs) in spherical harmonic (SH) domain has been thoroughly studied as a method of obtaining continuity over space. However, HRTFs are functions not only of direction but also of frequency. This paper presents an extension of the SH-based method, utilizing hyperspherical harmonics (HSHs) to obtain an HRTF representation that is continuous over both space and frequency. The application of the HSH approximation results in a relatively small set of coefficients which can be decoded into HRTF values at any direction and frequency. The paper discusses results obtained by applying the method to magnitude spectra extracted from exemplary HRTF measurements. The HRTF representations based on SHs and HSHs exhibit similar reproduction accuracy, with the latter one featuring continuity over both space and frequency and requiring much lower number of coefficients. The developed HSH-based continuous functional model can serve multiple purposes, such as interpolation, compression or parametrization for machine-learning applications.
\end{abstract}

\section{Introduction}
\label{s:introduction}

Continuous development of virtual and augmented reality applications rises need for efficient binaural audio processing algorithms. Especially important role in authentic recreation of an auditory scene is played by the directivity of human ears, which is commonly called head-related transfer functions (HRTFs). HRTFs are different for each individual as they depend on the shape of torso, head and pinna. Even though there are some similarities among sets of HRTFs obtained for different people, application of individually measured or matched HRTFs has been proven to improve the localization abilities \cite{Wenzel1993,Begault2001}. The importance of HRTFs is especially prominent in recognizing the position of a sound source placed in sagittal planes, where simple binaural cues such as interaural level difference (ILD) and interaural time difference (ITD) are mostly invariant \cite{Macpherson2002,Agterberg2012}.

HRTFs can be physically measured in two ways. Most measurements are performed by putting a pair of microphones inside the subject's ears and recording the response to the sound coming from different directions (e.g. \cite{Andreopoulou2015a,Zhang2012}). Alternatively, one can make use of Helmholtz reciprocity principle and place microspeakers inside the ears while the microphones are set around the subject \cite{Zotkin2006}. Either way, the results of the measurements are sets of HRTFs for a finite number of directions. They are usually stored in the form of discrete head-related impulse responses (HRIRs), for example using the Spatially Oriented Format for Acoustics (SOFA) \cite{SOFA,Majdak2013}.

Since HRTFs are relatively large data sets, some attempts were made to develop an efficient model that would reduce the amount of data without a significant loss of accuracy. Initially, the research concerned approximating HRIRs, e.g. by expressing them as filters of either finite or infinite impulse response  (e.g. \cite{Kulkarni1995,Kulkarni2004}), but soon the focus was moved to the spatial properties. In 1998, Evans et al. proposed expressing HRTFs by means of spherical harmonics (SHs) \cite{Evans1998}. Beside reducing the data size, this representation introduced an even more significant feature -- continuity over space. Over the years, the SH-based method has been widely investigated, regarding i.a. efficiency of different sampling schemes \cite{Zhang2012}, preprocessing techniques \cite{Brinkmann2018} or mixed-order approximations \cite{Ben-Hur2019}. Some alternative continuous representations were suggested, e.g. based on spherical wavelets \cite{Hu2019,Liu2019} or Slepian functions \cite{Bates2015}; however, SHs still prevail as the most popular basis functions to approximate not only HRTFs but also other directivity functions such as the sound source directivity \cite{Szwajcowski2021,Shabtai2017} or microphone directivity \cite{Ziegler2017}.

Beside focusing on either only spatial or only time/frequency variations, several attempts were made to develop a model including both these dependences. In 1992, Kistler and Wightman employed principal component analysis to an HRTF database \cite{Kistler1992}. While resulting model indeed provided good accuracy, it was discrete and required a priori knowledge of the database to determine optimal basis vectors. In 1995, Chen et al. developed a representation based on thin-plate splines that was continuous in space but still discrete in frequency \cite{Chen1995}. In 2009, Zhang et al. described a model based on Fourier series and Bessel functions which was continuous over both frequency and horizontal angle \cite{Zhang2009} and which was later extended to cover the entire sphere and include distance dependence as well \cite{Zhang2010}. Another proposition comes from Shekarchi et al. who based their model on the fusion of infinite impulse response filters and Legendre polynomials \cite{Shekarchi2013}. However, the authors focused purely on the compression of measured data and thus their model cannot be used to interpolate missing values. In 2015, Zhang et al. created a fully continuous functional model utilizing SHs for spatial dependency and complex exponentials for frequency representation \cite{Zhang2015}\footnote{Please note, that \cite{Zhang2015} was lead by Mengqiu Zhang, while \cite{Zhang2009} and \cite{Zhang2010} were lead by Wen Zhang.}. Both mentioned fully continuous models (\cite{Zhang2010} and \cite{Zhang2015}) determined coefficients for basis functions partially by numerical integrating, which is computationally efficient, but provides worse accuracy than that obtained by fitting in the least-squares sense. Furthermore, these researches focused on retrieving high accuracy complex representations, even though it is known that humans are insensitive to fine spectral or spatial details of HRTFs \cite{Kulkarni1998,Romigh2015,Breebaart2001,Xie2010}.

In this paper, a new approach to HRTF dimensionality is presented: frequency is imagined as another spatial dimension. This way, the character of these functions becomes purely spatial with two truly spatial variables (horizontal and vertical angles) and one extra variable that represents frequency, but is also treated as a spatial dependence\footnote{This paper concerns far-field HRTFs, where the radius dependence is neglected. This is why three- and four-dimensional spaces are described by only two and three variables, respectively.}. This approach emphasizes coupling between frequency and space in directivity functions and is inspired by the mathematical structure known as Minkowski space, where time and three-dimensional (3D) space are combined together to conveniently express some of the physical phenomena, most notably regarding theory of special relativity. Assuming that SHs are a good choice for expressing directivity functions in the 3D space, including the fourth dimension would require extention of the basis to a four-dimensional (4D) space. A similar problem was tackled in computer graphics, where standard 3D shape descriptors were insufficient for certain cases. In 2007, Bonvallet et al. extended the popular approach to another dimension by replacing SHs with hyperspherical harmonics (HSHs) \cite{Bonvallet2007}. This method was later applied in medical imaging by Pasha Hosseinbor et al. \cite{PashaHosseinbor2015}, providing further proof that HSHs, previously reserved for theoretical chemistry and nuclear physics, can be successfully employed in engineering. However, to the best of our knowledge, HSHs have not yet been utilized in acoustics or any related field. Thus, the application of this basis to represent acoustical data is the main novel element of this paper.

The primary advantage of proposed HSH representation over the state-of-the-art SH one is continuity over both space and frequency. Such representation allows to extract HRTF magnitude not only at a given direction but also at a given frequency without any additional operations (e.g. interpolation or resampling), thus being computationally attractive. Furthermore, varying the approximation parameters enables easy control of the balance between accuracy and amount of data; acknowledging psychoacoustical aspects in the process of deriving the functional model can lead to significant reduction of data size by ignoring high-order HSHs responsible for imperceptible spectral details. Last but not least, more holistic HRTF representation can be of great value in research requiring a thorough directivity parametrization method, e.g. for machine learning applications.

Section \ref{s:theory} provides necessary theoretical background on 4D coordinate systems and HSHs. Section \ref{s:hsha} describes conversion from raw measurement data to hyperspherical data and then to HSH domain. Section \ref{s:example} presents exemplary results of the conversion performed on a typical set of HRTFs and suggests further improvements of the process. Section \ref{s:discussion} consists of general comments on the HSH representation, its accuracy and potential applications. Finally, section \ref{s:conclusions} summarizes all the content of this paper.

\section{Theoretical background}
\label{s:theory}

HRTFs are functions of direction, distance and frequency. Under the assumption that the distance is greater than 1 m, the radial dependence can be dropped. Such simplified functions are called far-field HRTFs, but since they are measured and applied more commonly than full, distance-dependent HRTFs, they are often called just HRTFs. This is also the case in this paper; wherever HRTFs are mentioned, they mean far-field HRTFs, independent of distance.

\subsection{Hyperspherical coordinate system}
\label{s:coordinates}

Assuming that the spatial dependence of HRTFs is to be described in the spherical coordinate system, an extension of this system is needed to capture variability over another dimension. This can be achieved by adding another linear or angular dimension, resulting in either spherindrical or 4D hyperspherical coordinate system (HCS), respectively. The HCS has already proven to be successful in similar research problems in other fields \cite{Bonvallet2007,PashaHosseinbor2015} and its geometrical properties enhance natural physical properties of acoustic directivity characteristics (see Sec. \ref{s:poles} and \ref{s:mapping} for explanation). For these reasons, HCS was chosen over spherindrical coordinate system for this research.

HCS consists of hyperspherical radius $\rho$ and three angles: $\varphi \in [0,2\pi)$, $\theta \in [0,\pi)$ and $\psi \in [0,\pi)$. The angles $\varphi$ and $\theta$ correspond to azimuth and inclination defined in the spherical coordinate system, while $\psi$ is an extra angle representing another dimension. Since there are no unified  names for angles in 4D space, within this paper they are referred to just by their symbols or colloquially as the spatial angles (meaning angles $\varphi$ and $\theta$) and the frequency angle (meaning angle $\psi$)\footnote{The details of utilizing an angle to describe the frequency are provided in section \ref{s:mapping}.}. The relation between HCS and 4D Cartesian system ($x$,$y$,$z$,$w$) is following:

\begin{equation}
\begin{aligned}
	&x = \rho \sin{\psi} \sin{\theta} \sin{\varphi} \, ,\\
	&y = \rho \sin{\psi} \sin{\theta} \cos{\varphi} \, ,\\
	&z = \rho \sin{\psi} \cos{\theta} \, ,\\
	&w = \rho \cos{\psi} \, .
\label{e:HCS}
\end{aligned}
\end{equation}

\subsection{Hyperspherical harmonics definition}
\label{s:definition}

HSHs can be defined for any multidimensional space, but within this paper only 4D HSHs are considered and referred to simply as HSHs. They can be defined as following \cite{Domokos1967}:

\begin{equation}
	Z_{nl}^{m}(\varphi,\theta,\psi) \equiv N(n,l)\sin^l{\psi} \; C_{n-l}^{l+1}(\cos{\psi}) \; Y_{l}^{m}(\varphi,\theta) \, ,
\label{e:HSH}
\end{equation}

\noindent where $N(n,l)$ is a normalization factor, $C_{\nu}^{\alpha}(x)$ are Gegenbauer polynomials and $Y_{l}^{m}(\varphi,\theta)$ are SHs, while $n$, $l$ and $m$ are integer parameters limited as following:

\begin{equation}
\begin{aligned}
	&n \geq 0 \\
	0 \leq \, &l \leq n \\
	-l \leq \, &m \leq l \, .
\label{e:nlm}
\end{aligned}
\end{equation}

$N(n,l)$ is a normalization factor, making the HSH basis not only orthogonal but orthonormal. It is given by the formula:

\begin{equation}
	N(n,l) \equiv 2^{l+\frac{1}{2}} (l+1)! \sqrt{\frac{2(n+1)(n-l+1)!}{\pi (n+l+2)!}} \, .
\label{e:normalization}
\end{equation}

\noindent It is worth noting, that SHs in Eq. (\ref{e:HSH}) are also normalized, which is why $N(n,l)$ does not depend on the parameter $m$ - this dependency is entirely included in the SH normalization factor.

$C_{\nu}^{\alpha}(x)$ are Gegenbauer polynomials, also known as ultraspherical polynomials. They can be defined in multiple equivalent ways; however, for the sake of computations, the most useful one is that given by the following recurrence relation \cite{gegenbauer}:

\begin{equation}
\begin{aligned}
	C_{0}^{\alpha}(x) = &1 \\
	C_{1}^{\alpha}(x) = &2 \alpha x \\
	C_{\nu}^{\alpha}(x) = &\frac{1}{\nu} \big{(}2x(\nu+\alpha-1)C_{\nu-1}^{\alpha}(x) - (\nu+2\alpha-2)C_{\nu-2}^{\alpha}(x) \big{)} \,  .
\label{e:gegenbauer}
\end{aligned}
\end{equation}

\noindent In the HSH definition (Eq. (\ref{e:HSH})), Gegenbauer polynomials, together with the factor of $\sin^l{\psi}$, are responsible for variation among $\psi$ angle (the frequency angle).

$Y_{l}^{m}(\varphi,\theta)$ are SHs. SHs can be defined either as complex- or real-valued functions. This paper focuses on their application to express magnitude spectra of HRTFs (see section \ref{s:hrirs24D} for explanation) and thus the real form is used, making the entire HSH basis real as well. Real SHs are defined as following\footnote{Sometimes in real SH definitions there is also $(-1)^{m}$ factor included, but it is irrelevant for approximation purposes as it merely changes the sign of some coefficients.} \cite{Varshalovich1988}:

\begin{equation}
Y_{l}^{m}(\phi,\theta) \equiv
\begin{cases}
N_{Y}(l,m) P_{l}^{m}(\cos{\theta}) \cos{\left( m\varphi \right)} \, , & \text{if}\ m \geq 0 \\
N_{Y}(l,m) P_{l}^{|m|}(\cos{\theta}) \sin{\left( |m|\varphi \right)} \, , & \text{if}\ m < 0 \\
\end{cases} \, ,
\label{e:SH}
\end{equation}

\noindent where $P_{l}^{m}$ are the associated Legendre functions and $N_{Y}(l,m)$ is the normalization factor for SHs defined as:

\begin{equation}
 N_{Y}(l,m) \equiv \sqrt{(2-\delta_{m0})\frac{2l+1}{4 \pi} \frac{(l-|m|)!}{(l+|m|)!}} \, ,
\label{e:normalizationSH}
\end{equation}

\noindent where $\delta$ is Kronecker's delta function:

\begin{equation}
\delta_{ij} \equiv
	\begin{cases}
	1, & \text{if}\ i=j \\
	0, & \text{if}\ i \neq j \\
	\end{cases} \, .
\label{e:Kronecker}
\end{equation}

\subsubsection{Hyperspherical poles}
\label{s:poles}

Hyperspheres, by being extensions of 2-spheres (regular 3D spheres), display some analogous features. On a 2-sphere, there are two poles lying at inclinations $\theta \in \{ 0, \pi \}$, at which all meridians converge to the same point. In other words, at these poles, the actual direction is the same for every azimuthal angle $\varphi$. In HSC, such poles exist for every $\psi$, but there are also two major poles (hyperpoles) at $\psi \in \{ 0, \pi \}$ where the direction is dependent on neither $\varphi$ nor $\theta$. Now, applying the interpretation that $\psi$ is the frequency angle, this means that values for the limit frequencies have to be constant for every physical direction. This is inline with how directivity (including HRTFs) behaves at low frequency - the lower the frequency, the less variance along the directions can be observed, as the size and shape of objects becomes less and less relevant when compared to the length of acoustic waves corresponding to the analyzed frequency. This coincidence is taken advantage of in the mapping of frequency to angle, so that the hyperpole convergence can be perceived as an advantage rather than a limitation of the proposed model (see Sec. \ref{s:mapping}).

\section{Hyperspherical harmonic approximation}
\label{s:hsha}

In order to express HRTFs in the HSH domain, two steps have to be taken: first, the measurement data have to be converted to HCS and then the HSH coefficients need to be determined so that the weighted sum of HSHs approximates the data in HCS as accurately as possible. Both these steps are described in details in the following subsections.

\subsection{HRIRs to data in HCS}
\label{s:hrirs24D}

As stated in Introduction, HRTFs are commonly stored in the form of HRIRs. It might seem natural to try to find a way to efficiently represent the data in time domain rather than in the frequency domain. However, there are several arguments in favour of the latter. First of all, in previous research, both approaches were studied and models based on the frequency representation resulted in lower approximation errors using objective measures \cite{Evans1998,Hartung1999}. Secondly, the human auditory system, analyzes sounds mostly in the frequency domain and thus accurate reproduction of the impulse response shape is irrelevant to our ears, as opposed to the shape of the frequency spectrum. Furthermore, HRIRs usually have much sharper shapes than the corresponding HRTFs, which makes representing them as a sum of basis functions much harder, especially when the basis is truncated. Last but not least, HRIRs include information on phase, which is of little importance; it is widely acknowledged that phase spectra can be ignored as long as the interaural time difference is preserved \cite{Kistler1992,Romigh2015,Kulkarni1999}\footnote{Phase spectra are still important for low frequencies, but they can be replaced by a linear phase derived from interaural time differences, without any noticeable damage to the localization abilities.}, although some contrary results have also been presented \cite{Rasumow2014}. Basing a model on the frequency-domain representation thus allows the information on phase to be dropped, reducing the efficient amount of data twice without a significant loss as far as the applicatory aspect is concerned. However, HSHs can be defined as complex functions as well, so it is possible to include the phase information as well, if desired.

The magnitude spectra typically are represented either in linear or in logarithmic scale (in dBs). Referring again to the psychoacoustic aspects of sound perception, the logarithmic scale is relative and thus it is more reflective of how the sound pressure is perceived by the human auditory system. Since the approximation method of the  model proposed within this paper is based on the least-squares fitting (see section \ref{s:computation} for more details), it seems reasonable to use logarithmic scale, so that the solver minimizes relative errors \cite{Kulkarni2004,Romigh2015,Hartung1999,Blommer1997}.

The reference value for the logarithmic scale does not matter as long as it is uniform within a given HRTF set; this value is linked only with the coefficient for $Z_{00}^{0}$ (first HSH invariant along all the angles), but otherwise has no impact whatsoever. Therefore, for simplicity sake, the reference value was chosen to be 1.

\subsubsection{Frequency mapping}
\label{s:mapping}

\begin{figure}[ht]
\centering
\includegraphics[width=\linewidth]{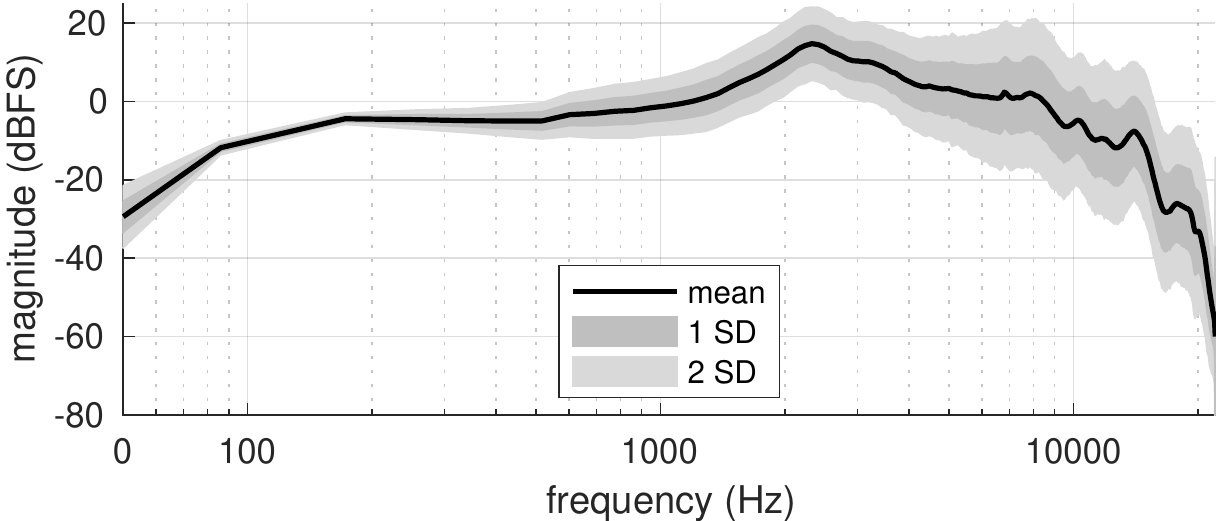}
\caption{Statistical distribution of an exemplary set of HRTFs (described in details in section \ref{s:example}) for all directions. Solid black line is the mean spectrum, while the dark and light grey areas cover regions of 1 and 2 standard deviations, respectively. First frequency bin was shifted from 0 to 50 Hz, to be able to fit on the logarithmic axis. At the last frequency bin, standard deviation is especially large due to the occurrence of zero points that were converted to floating-point accuracy in MATLAB to avoid $-\infty$ values. The plot was shifted to fluctuate around 0 dB.}
\label{f:dist}
\end{figure}

As signaled earlier, in order to express HRTFs in HCS, frequency has to be mapped to the $\psi$ angle. The simplest way of doing it would be to set the frequency $f = 0$ at $\psi = 0$ and the frequency $f = f_{s}/2$ to $\psi = \pi$, where $f_{s}$ denotes sampling frequency of HRIRs in the HRTF set. The remaining frequencies would be then mapped to $\psi$ linearly, following the formula:

\begin{equation}
\psi_{k} = \frac{2 \pi f_{k}}{f_{s}} \, ,
\label{e:mapping_prev}
\end{equation}

\noindent where $f_{k}$ is the center frequency of $k$th frequency bin and $\psi_{k}$ is the corresponding value of the frequency angle. However, such mapping implies convergence to uniform values at hyperpoles, i.e. at $f \in \{0, f_{s}/2\}$. When approaching 0 Hz, HRTFs tend to be omnidirectional anyway, but this is not the case for the highest frequencies, especially in the logarithmic scale (see Fig. \ref{f:dist}). While this might seem to be of little importance, as the convergence to the hyperspherical pole at the Nyquist frequency would likely affect mostly the highest frequencies which are inaudible anyway, it is possible to avoid such distortions. $f = f_{s}/2$ can be set to $\psi = \pi/2$ instead of $\psi = \pi$, so that all the HRTFs can be mapped to effectively only half a hypersphere. This requires twice as large resolution of HSH along $\psi$ since the spectra are squeezed to fit on only half of the full range of the frequency angle; however, such mapping allows to ignore all HSHs that are not symmetric about the hyperplane at $\psi = \pi/2$. Application of only $\psi$-symmetric HSHs (the HSHs that exhibit said symmetry about $\psi = \pi/2$) removes the convergence at $f = f_{s}/2$ and makes the spectra for the entire range of $\psi \in [0,\pi)$ symmetric about the Nyquist frequency in the same manner as typical magnitude spectra obtained by performing discrete Fourier transform on a real signal. With $f = f_{s}/2$ set to $\psi = \pi/2$, the Eq. (\ref{e:mapping_prev}) takes the form of:

\begin{equation}
\psi_{k} = \frac{\pi f_{k}}{f_{s}} \, .
\label{e:mapping}
\end{equation}

Alternatively, different frequency mappings could be used. Both limit values and the character of scale can be changed, e.g. to logarithmic frequency scale, which is quite popular in audio engineering, or the mel scale, which is based on a psychoacoustic model of frequency perception. This paper concerns only the linear frequency mapping and leaves the employment of other scales for potential future research.

\subsection{Data in HCS to HSH coefficients}
\label{s:4D2hsh}

Once the 4D data are prepared in the form of a set of triplets of angles $(\varphi,\theta,\psi)$ and the corresponding magnitudes, these data need to be approximated by a finite number of HSHs. The approximation comes down to determining the coefficients $\alpha_{n,l}^{m}$ for the weighted sum so that approximated function $\hat{H}(\varphi,\theta,\psi)$ of the form\footnote{Equation (\ref{e:HSHsum}) assumes approximation using all the HSHs up to $n = n_{text{max}}$. However, all of the parameters $n$, $l$ and $m$ can have independent limits (see section \ref{s:parameters})}:

\begin{equation}
\hat{H}(\varphi,\theta,\psi) = \sum_{n=0}^{n_{\text{max}}} \sum_{l=0}^{n} \sum_{m=-l}^{l} \alpha_{nl}^{m} Z_{nl}^{m}(\varphi,\theta,\psi)
\label{e:HSHsum}
\end{equation}

\noindent is as close to the original function $H(\varphi,\theta,\psi)$ as possible. In the analytical approach, when an infinite number of basis functions can be used ($n_{\text{max}} = \infty$), consecutive coefficients can be determined by calculating the dot product of the approximated function and complex conjugation of the given HSH:

\begin{equation}
\alpha_{n,l}^{m} = \int_{\mathbb{S}^{3}} H(\varphi,\theta,\psi)  \overline{Z_{nl}^{m}(\varphi,\theta,\psi)} d \chi \, ,
\label{e:dotproduct}
\end{equation}

\noindent where $\int_{\mathbb{S}^{3}} d \chi$ means integrating over the surface of the unit 3-sphere:

\begin{equation}
\int_{\mathbb{S}^{3}} d \chi = \int_{0}^{2 \pi} \int_{0}^{\pi} \int_{0}^{\pi} \sin^{2}{\psi} \, d \psi \sin{\theta} \, d \theta \, d \varphi \, .
\label{int3D}
\end{equation}

\noindent However, when the number of basis functions is truncated, the accuracy of the approximation decreases. For this reason, least-squares fitting of the coefficient values is preferred over the numerical integration of Eq. (\ref{e:dotproduct}) to find the optimal set of coefficients.

\subsubsection{Determination of the number of HSHs used}
\label{s:parameters}

The choice of the number of basis functions is one of the most important decisions impacting the quality of the approximation. In general, increasing this number improves the overall accuracy of the approximation. However, using too many basis functions requires more computational resources and can lead to overfitting. It is thus important to determine what is the optimal number of basis functions for a given problem.

HSHs are described by three parameters: $n$, $l$ and $m$. Analyzing the definition given in Eq. (\ref{e:HSH}), one can notice that these parameters are responsible for variations along the angles $\psi$, $\theta$ and $\varphi$, respectively. Limiting just $n$ results in the same spatial frequency along all the angles. However, some extra limitations for $l$ and $m$ can be imposed to obtain different resolution along angles $\theta$ and $\varphi$. Equation (\ref{e:HSHsum}) then takes the form of:

\begin{equation}
\begin{aligned}
\hat{H}(\varphi,\theta,\psi) =  \sum_{n=0}^{n_{\text{max}}} \quad \sum_{l=0}^{\min{(n,l_{\text{max}})}} \sum_{m=-\min{(l,m_{\text{max}})}}^{\min{(l,m_{\text{max}})}} \alpha_{nl}^{m} Z_{nl}^{m}(\varphi,\theta,\psi) \, ,
\end{aligned}
\label{e:HSHsumLimits}
\end{equation}

\noindent where the limits have to satisfy the following relation\footnote{It is possible to obtain higher spatial frequency e.g. along $\theta$ than along $\psi$ by using only a selection of HSHs for higher values of the corresponding parameters. However, this case is not applicable within this paper and thus it is not further discussed.}:

\begin{equation}
n_{\text{max}} \geq l_{\text{max}} \geq m_{\text{max}} \, .
\label{e:nlmlimits}
\end{equation}

These maximum values of $n_{\text{max}}$, $l_{\text{max}}$ and $m_{\text{max}}$ should be determined with respect to the sampling theorem and depend on both spatial and frequency sampling. However, HRTFs do not need to be reconstructed with the highest available accuracy, since fine details are psychoacoustically irrelevant. Thus, it is important to find the minimum values of the limits for which the approximation error is negligible.

Spatial dependence is described by SHs and thus the limits for spatial frequency ($l_{\text{max}}$ and $m_{\text{max}}$ ) can be determined basing on SH-related research. Romigh et al. noticed that some of the localization abilities occur for HRTFs approximated by SHs of maximum order as low as 2 \cite{Romigh2015}. Approximations using SHs of maximum order 6 are already statistically indistinguishable from the unprocessed HRTFs and the localization does not improve with further increasing the maximum order. It is important to notice that Romigh et al. used the same preprocessing as proposed within this paper, i.e. least-squares fitting based on magnitudes in logarithmic scale. These results suggest that setting $l_{\text{max}}$ and $m_{\text{max}}$ to 6 should be sufficient to prevail the localization properties of a set of HRTFs.

Similar research was conducted regarding sensitivity to spectral details of HRTFs. Kulkarni and Colburn checked how much smoothing in frequency can be applied to HRTFs without a significant impact on the localization abilities \cite{Kulkarni1998}. For this reason, they prepared stimuli by truncating the Fourier series representing the logarithmic spectra of empirical HRTFs. The listeners that took part in the experiment were unable to discriminate between the real and smoothed virtual sound sources in most of the setups for HRTFs reconstructed from only 16 coefficients. For 32 coefficients and above, all of the listeners performed at chance for all the tested directions. Even though the frequency angle in HSHs is not represented by the Fourier series, the same spatial frequency along that angle can be chosen by setting approppriate $n_{\text{max}}$. Since for chosen frequency mapping the spectra are supposed to fit on only half the hypersphere, $n_{\text{max}}$ must be doubled to exhibit desired resolution over the effectively used range of $\psi$. Thus, to match the spatial frequency of 32 spectrum coefficients from the described experiment, $n_{\text{max}}$ should be set to 64.

Furthermore, as stated in section \ref{s:mapping}, effectively utilizing only half of the hypersphere allows to ignore non-$\psi$-symmetric HSHs. Following the HSH and Gegenbauer polynomials definitions given in Eqs (\ref{e:HSH}) and (\ref{e:gegenbauer}), it can be noted that for $Z_{nl}^m$ to be $\psi$-symmetric, the difference between $n$ and $l$ has to be even. All the parameters configurations where this difference is odd should be thus disregarded when summing over $l$ in Eq. (\ref{e:HSHsumLimits}).

\subsubsection{Computations}
\label{s:computation}

Assuming sampling at $K$ 4D directions ${\Omega_{k}\equiv (\varphi_{k}, \theta_{k}, \psi_{k})}$, the HSH coefficients can be determined by solving the following matrix equation:

\begin{equation}
\label{e:matrixHSH}
\begin{bmatrix}
Z_{00}^{0}(\Omega_{1}) & \dots & Z_{n_{\text{max}}l_{\text{max}}}^{m_{\text{max}}}(\Omega_{1}) \\
\vdots & \ddots & \vdots \\
Z_{00}^{0}(\Omega_{K}) & \dots & Z_{n_{\text{max}}l_{\text{max}}}^{m_{\text{max}}}(\Omega_{K}) \\
\end{bmatrix}
\begin{bmatrix}
\alpha_{00}^{0} \\ \vdots \\ \alpha_{n_{\text{max}}l_{\text{max}}}^{m_{\text{max}}}
\end{bmatrix} \\
 =
\begin{bmatrix}
H(\Omega_{1}) \\ \vdots \\ H(\Omega_{K})
\end{bmatrix} \, ,
\end{equation}
\vspace{5mm}

\noindent where $H(\Omega)$ are given in dBs to minimize errors in logarithmic scale.

Since the system is overdetermined, it is usually impossible to find coefficients which would perfectly satisfy this equation. However, it can be solved in the least-squares sense, minimizing the error of approximation. This approach is commonly embraced in research concerning SH approximation \cite{Zhang2012,Romigh2015,Pasqual2014,Alon2018}. A few existing 4D directivity models featured separate computation of spatial and frequency dependences, using least-squares fitting for space and then direct integration for frequency \cite{Zhang2010,Zhang2015}. Including both frequency and spatial dependent functions in a single least-squares minimization to the best of my knowledge has not been yet applied for these kind of data, making it another novel element of the paper. While such approach demands more computational resources, it provides better fitting and acknowledges coupling of space and frequency in directivity functions. It is also worth noting that the increased computational complexity mainly concerns determining the HSH coefficients and is caused by replacing multiple smaller matrix equations by one large. In real-time, with proper optimization, rendering binaural sound should be comparably fast for both SH and HSH HRTF representations.

HRTF sets usually lack data for low elevation angles ($\theta \rightarrow \pi$) because of the measurement setup restrictions. This can lead to some irrational values in the unsampled region, since the least-squares solver minimizes error only at the points where data is available. One way of dealing with this effect is to apply a proper regularization \cite{Zhang2015,Zotkin2009}. Ahrens et al. proposed another, even more efficient solution based on filling the missing region by means of low-order SH approximation \cite{Ahrens2012}. However, since in HSH definition the direction dependence is described by SHs, it is logical to assume that these issues can be handled for the HSH approximation in the same manner.  For the sake of comparison between HSHs and SHs, applying any of these methods should not impact the analysis results (both approximations are expected to be impacted in the same way). Thus, simple, non-regularized least-squares fitting was used for both bases.

The computations and analysis were performed in \textsc{Matlab} using \textsc{ooDir} Toolbox \cite{ooDir}. Both the classes used in this research as well as precomputed objects containing raw and approximated data can be found in its database.

\section{Exemplary approximations}
\label{s:example}

To showcase efficiency of the HSH approximation, the method was tested on exemplary data. The chosen set of HRTFs comes from the original measurements of KEMAR (Knowles Electronics Manikin for Acoustic Research) with large pinnae performed at Massachusetts Institute of Technology \cite{Gardner1995} and is a typical HRTF set for evaluating different models \cite{Zhang2009,Zhang2010,Shekarchi2013,Zhang2015}. The set contains 710 HRIRs for each ear, measured at different directions and each consisting of 512 samples recorded with the sampling frequency of 44.1 kHz. The HRIRs were converted to HCS as described in section \ref{s:hrirs24D}. Only data for the left ear was used. 

Although, according to section \ref{s:parameters}, the parameter limits of $\{l_{\text{max}}, m_{\text{max}}\} = 6$ and $n_{\text{max}} = 64$ should be sufficient, it is important to remember that the SH-focused research used exact frequency spectra, and the spectral-focused one used exact spatial representation. Smoothing in both space and frequency can thus have more impact on the localization abilities. For this reason, the limits used in the exemplary approximation were set to slightly higher values of $\{l_{\text{max}}, m_{\text{max}}\} = 8$ and $n_{\text{max}} = 80$.

\begin{figure}[ht!]
\centering
\subfigure[]{
	\includegraphics[width=\linewidth]{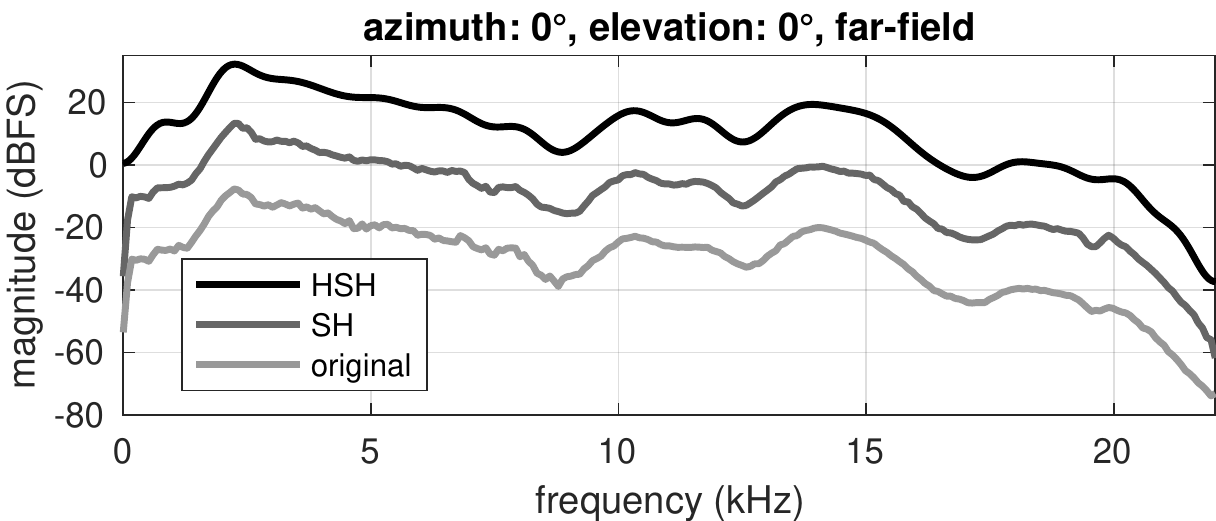}
	\label{f:ex00}}
\\
\subfigure[]{
	\includegraphics[width=\linewidth]{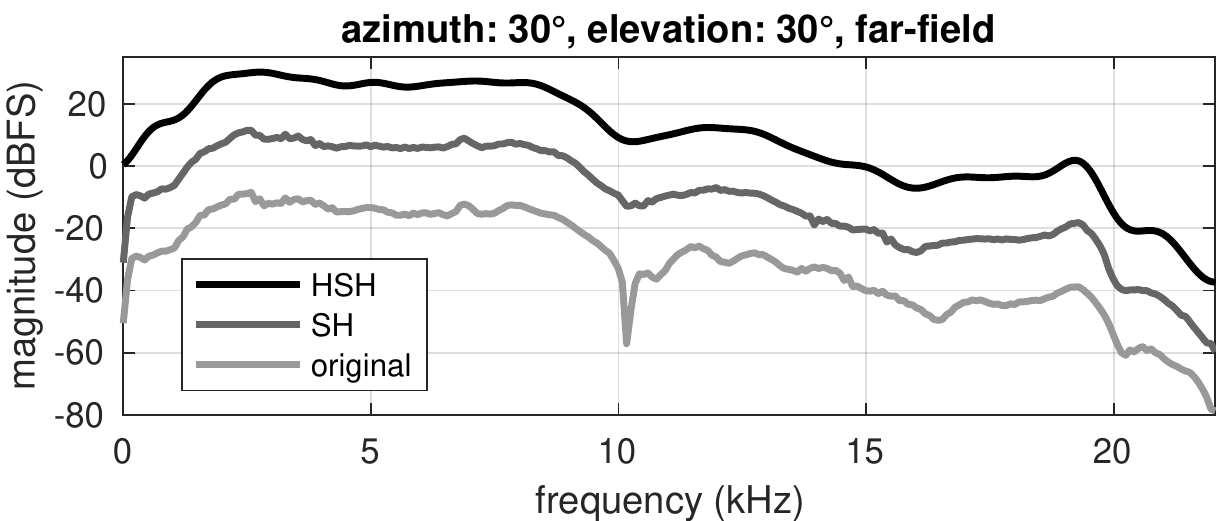}
	\label{f:ex3030}}
\caption{Raw and approximated spectra of KEMAR HRTFs for two exemplary directions: a) azimuth: 0\degree{}, elevation: 0\degree{}; b) azimuth: 30\degree{}, elevation: 30\degree{}. The spectra are offset by 20 dB for clarity. See text for more details.} 
\label{f:example}
\end{figure}

For the sake of comparison, the SH approximation was also performed on the same exemplary data. The procedure was exactly the same as for the HSH approximation, but the computations were carried out separately for each frequency bin. Then, approximated values for given directions were put together to retrieve discrete spectra. Maximum order and degree of SHs used for approximation was set to 8 to match the limiting parameters of SHs embedded in the HSHs. Exemplary raw spectra and their SH and HSH approximations are shown in Fig. \ref{f:example}.

The HSH approximation retrieves the general shape very well, but the resulting spectra are smoother than both the ones computed from the raw data and the ones obtained by means of the SH approximation. The most problematic details to represent in the HSH domain are deep notches, e.g. around 10 kHz in Fig. \ref{f:ex3030}. However, this is not exclusive to HSH approximation but comes from the spatial dependence represented by SHs, as the SH approximation features the same issues. The notches are caused by rapid changes in phase. Zagala and Zotter developed a method to deal with such a problem for low-order SHs \cite{Zagala2019}. Their algorithm could be extended and applied to HSHs if necessary.

By the properties of $\psi$-symmetric HSHs, their first derivative with respect to $\psi$ has to be 0 at $\pi/2$, which causes minor discrepancies at the highest frequencies. However, since they lie beyond human hearing range, these discrepancies are irrelevant from the perceptual point of view. Another problematic frequency band is the very lowest one. In this case, the SH approximation is much more precise than the HSH one; for low frequencies, the directivity takes very simple shapes (mostly omnidirectional), which are easy to express in the SH domain. One of the reasons for the inaccuracy of the HSH approximation in this band are rapid changes of magnitude levels within first few frequency bins, which come from the audio chain used in the HRTF measurement rather than the physical properties of HRTF themselves. This issue is investigated in depth in the following subsection.

\subsection{Frequency weighting}
\label{s:weighting}

Since some frequency bands lie beyond human hearing range, their impact on the approximation results should not be as high as the impact of the audible part of the spectra. This variation in relevance can be achieved by applying weighting to the least-squares minimization. The Eq. (\ref{e:matrixHSH}) then takes the form of:

\begin{equation}
\label{e:weightedLS}
(\mathbf{Z^{\mathrm{T}} W Z) \bm{\upalpha} = Z^{\mathrm{T}} W H} \, ,
\end{equation}

\noindent where $\mathbf{Z}$, $\bm{\upalpha}$, and $\mathbf{H}$ denote the respective matrices from Eq. (\ref{e:matrixHSH}), $^{\mathrm{T}}$ denotes matrix transposition and $\mathbf{W}$ is a diagonal matrix with weights for consecutive angle triplets:

\begin{equation}
\label{e:weightMatrix}
\mathbf{W} =
\begin{bmatrix}
w(\Omega_{1}) & 0 & \dots & 0 \\
0 & w(\Omega_{2}) & & \vdots \\
\vdots & & \ddots & 0 \\
0 & \dots & 0 & w(\Omega_{K}) \\
\end{bmatrix} \, .
\end{equation}
\vspace{5mm}

One of the regions where fitting could be improved by applying proper weights is the frequency band represented by the first few frequency bins. For all the directions, a drop-off can be noticed as the frequency approaches 0, which could be caused by low-frequency limit of sound sources used in the HRTF measurement or a high-pass filter embedded in the audio chain. For the lowest frequencies, where the corresponding sound waves are much longer than the size of a human head, the HRTFs can be considered omnidirectional (see Fig. \ref{f:dist}). This omnidirectionality mirrors the convergence of HCS for the frequency angle $\psi \rightarrow 0$. Since the problematic first bins do not hold any important information, they can be completely removed at the stage of computing HSH approximation (i.e. have weight equal to 0). However, one needs to be wary that removing data from certain regions can under some circumstances lead to overfitting.

The other perceptually irrelevant part of the spectra are the highest frequencies. In this case, there are more than one frequency bins lying beyond the human hearing range. Their weights were arbitrarily chosen to start decreasing above 20 kHz and reach 0 at the Nyquist frequency following the shape of cosine function.

\begin{figure}[h!]
\centering

\includegraphics[width=\linewidth]{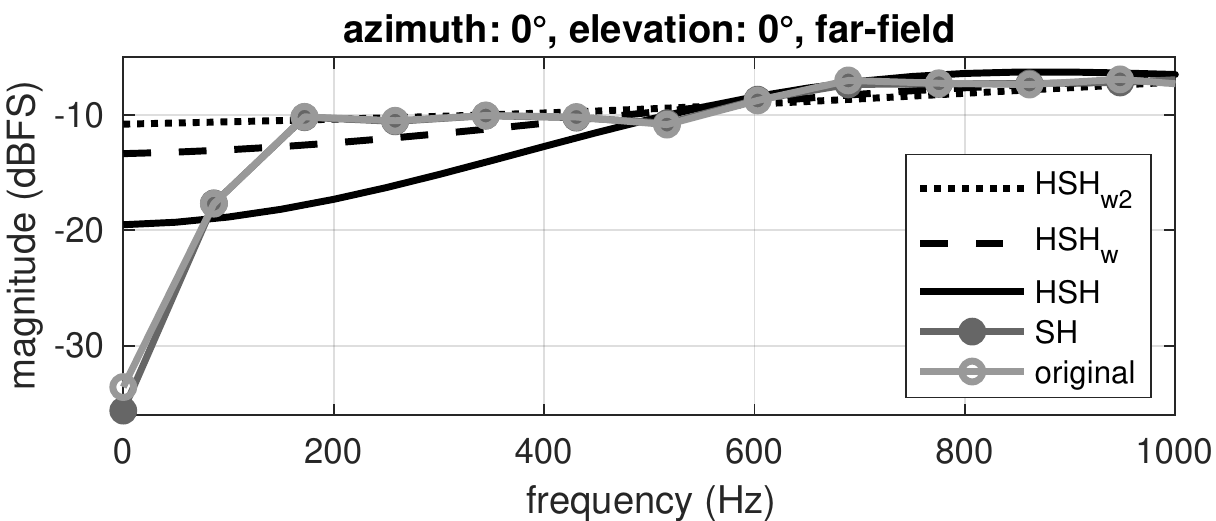}
\caption{Raw and approximated spectra of KEMAR HRTF for the direction straight ahead zoomed on the lowest frequencies. HSH$_w$ and HSH$_{w2}$ denote the HSH approximation with weights applied ignoring first or first two frequency bins, respectively (see text for more details).} 
\label{f:example_f0h_zoom}
\end{figure}

The HSH approximation was performed once again, applying the weights as described above. For most parts of the frequency spectrum, no noticeable changes occurred in regard to the previously performed HSH computation on the complete data. Since there is on average more weight on the audible frequency band, the resulting approximation should be closer to the original data, but the improvements were minuscule in most of the frequency range. However, the fitting for the lowest frequencies has been notably improved, except for the region corresponding to the first bins, which were effectively removed from the least-squares minimization (Fig. \ref{f:example_f0h_zoom}). Even though the resulting approximation is less accurate with regard to the measured data (large errors at first two bins), it is likely more accurate with regard to the factual physical HRTFs and improves fitting at several following bins.

From the perceptual point of view, the fitting in regions outside of the hearing range is irrelevant, and so the reduced impact of extreme frequency bins should improve the overall accuracy, even if only slightly. Thus, the following sections consider only this new approximation.

\subsection{Approximation error analysis}
\label{s:errors}

\begin{figure}[h]
\centering
\includegraphics[width=\linewidth]{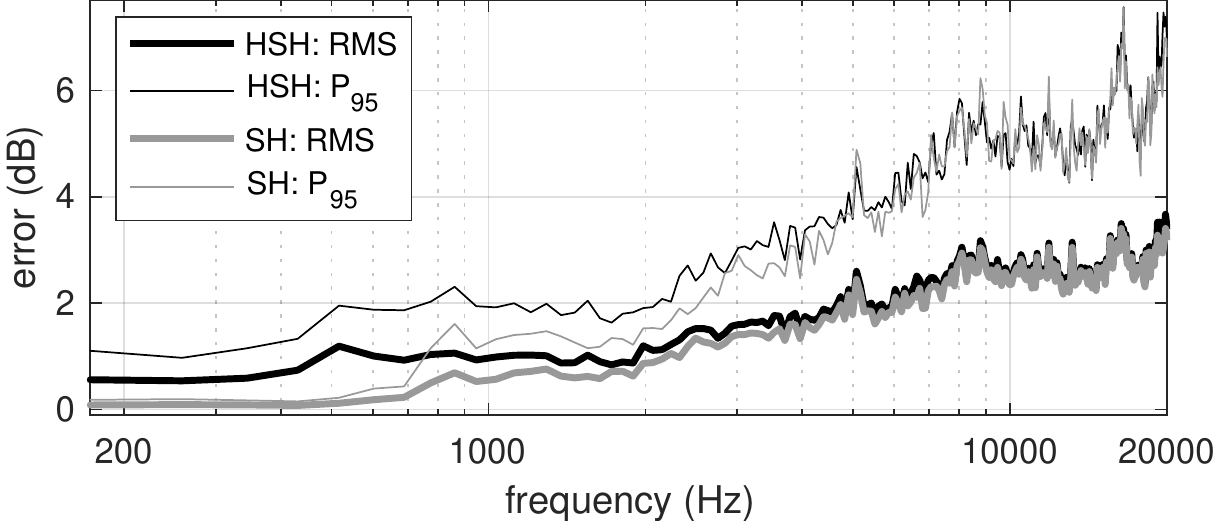}
\caption{RMS and P$_{95}$ of the HSH and SH approximation errors in reference to the discrete data}
\label{f:RMS}
\end{figure}

\begin{figure}[h]
\centering
\includegraphics[width=\linewidth]{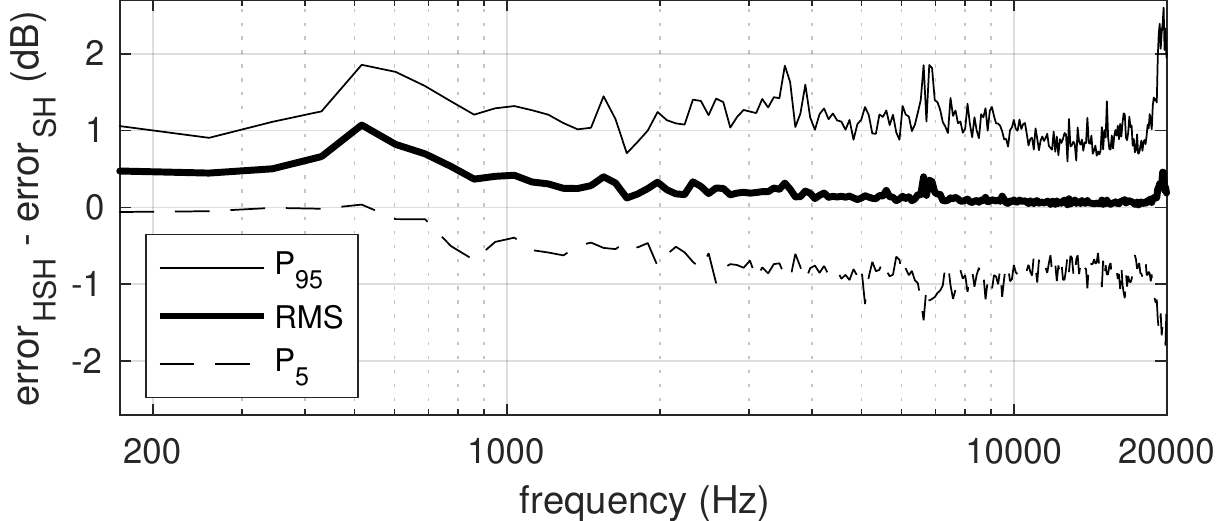}
\caption{Difference in RMS errors and 5th and 95th percentile of differences between the HSH and SH approximation errors}
\label{f:RMSdiff}
\end{figure}

Since 4D data is difficult to display efficiently as a 2D image, it is required to introduce numerical measures to compare accuracy of the SH and HSH approximations over the whole sphere. The key difference between these two approaches is the way of handling the frequency dependence and thus, for the sake of comparison, the approximation error should be presented in the frequency domain. In similar research, various measures of approximation error were used. Currently, one of the most popular approaches is to evaluate it as root-mean-square (RMS) error of the differences between logarithmic spectra (e.g. \cite{Liu2019,Romigh2015,Nishino1999,Li2021}). A common alternative is to evaluate the error based on the linear magnitude (e.g. \cite{Hu2019,Zhang2015}). As of the time of writing this paper, there are no clear indications on one of these approaches being superior to the other other than authors' speculations. However, the measure based on dB is more intuitive to interpret and the computations were performed on the logarithmic data, so it seems appropriate to use the logarithmic error as well. RMS errors for the HSH approximation\footnote{From this section on, the HSH approximation is the one with first two frequency bins removed and weights applied for the high frequencies (earlier denoted as HSH$_{w2}$).} were computed by averaging over all 710 directions available in the measurement data using the following formula:

\begin{equation}
\text{RMS}_{\hat{H}}(f)  = \sqrt{ \frac{ \sum_{k = 1}^{K} \left(\hat{H}(\varphi_{k},\theta_{k},\psi_{k}) - H(\varphi_{k},\theta_{k},\psi_{k})\right)^{2 } \delta_{\psi_{k}\psi_{f}}}{ \sum_{k = 1}^{K} \delta_{\psi_{k}\psi_{f}}} } \, ,
\label{e:RMS}
\end{equation}

\noindent where $\psi_{f}$ is the frequency angle corresponding to the frequency $f$, derived from Eq. (\ref{e:mapping}). RMS error for SH approximation was computed in an analogous way. In addition, for both methods and for each frequency bin, the 95th percentile (P$_{95}$) was determined for absolute values of $\hat{H}(\Omega)-H(\Omega)$. Errors in the form of RMS an P$_{95}$ values for both the SH and HSH approximations are plotted in Fig. \ref{f:RMS}, while Fig. \ref{f:RMSdiff} shows difference between the RMS plots from Fig. \ref{f:RMS} as well as 5th and 95th percentile determined on the sets of differences between absolute errors in dB for these two methods:.

\begin{equation}
\text{P}_{x \,_{\text{HSH-SH}}}(f) = \text{P}_{x}(\{|\hat{\mathbf{H}}^{f}_{\text{HSH}} - \mathbf{H}^{f}| - |\hat{\mathbf{H}}^{f}_{\text{SH}} - \mathbf{H}^{f}|\}) \, ,
\label{e:P95}
\end{equation}

\noindent where $\mathbf{H}^{f}$, $\hat{\mathbf{H}}^{f}_{\text{SH}}$ and $\hat{\mathbf{H}}^{f}_{\text{HSH}}$ are sub-vectors of the original data, SH and HSH approximations, respectively, which contain only values for a given frequency $f$.

For the vast majority of the hearing range, the difference in accuracy between the HSH and SH approximations is small, reaching about 1 dB around 500 Hz and becoming less and less significant as the frequency rises. The largest discrepancies occur for the lower frequency regions (below 1 kHz), where SH approximation provides near perfect reconstruction of the original data, but even there the absolute error is still overall low (around 0.5-1 dB). The values of P$_{5}$ indicate that in some cases the HSH approximation yields lower errors than the SH one (especially for higher frequencies). The RMS error, however, is lower for SH approximation for each frequency bin, even if only ever so slightly.

\subsubsection{Single-term error}

To improve the comparability of the HSH approximation accuracy, the error needs to be provided as a single value averaged over all directions and frequencies. For logarithmic scale, an error measure called spectral distortion (SD) is usually employed. Its definition resembles the RMS definition in Eq. (\ref{e:RMS}) but without the Kronecker's deltas, so that the averaging is performed over both space and frequency:

\begin{equation}
\text{SD}_{\hat{H}}  = \sqrt{ \frac{1}{K} \sum_{k = 1}^{K} \left(\hat{H}(\Omega_{k}) - H(\Omega_{k})\right)^{2 } }\, .
\label{e:SD}
\end{equation}

SD values were computed for both SH and HSH approximations in the range 100 Hz to 20 kHz (the lower limit was imposed to exclude the first two frequency bins). The results were $\text{SD}_{\text{SH}} = 2.32$ dB and $\text{SD}_{\text{HSH}} = 2.44$ dB. Since SD requires a ground truth data, it cannot accurately capture the magnitude of differences between two approximations and so the difference between SD values is more representative than the SD of the differences between approximated data (similarly to difference of RMS being more accurate than RMS of differences in Fig. \ref{f:RMSdiff}).

\section{Discussion}
\label{s:discussion}

In general, the HSH and SH approximations provide very similar accuracy within the hearing range. However, it is important to note that the minuscule difference in the approximations' errors for middle and audible high frequencies does not mean that there are no errors within these bands. In fact, the errors in both cases are lower for the lowest frequencies, but the SH approximation is very accurate in this region and thus the difference is larger than for the middle and high frequencies. Such characteristics suggest that the lower accuracy of HSH approximation for higher frequencies comes from inability of low-order SHs to properly express the spatial variability of HRTFs in these bands. However, the maximum order of SHs was chosen based on perceptual studies and so it indicates that these magnitudes of errors are imperceptible. It is notable, that the approximations are not identical even when the RMS error is close to 0; the differences are small (mostly below 1 dB) and are distributed relatively evenly between being in favor of either SH or HSH approximation. Such small differences in middle and high frequencies have been proven to be perceptually insignificant by many researchers (e.g. \cite{Breebaart2001,Xie2010,Huopaniemi1999}).

At low frequencies, the accuracy of the HSH approximation improved significantly after removing the first two frequency bins from the least-squares minimization. However, the SH approximation still outperforms the HSH one, providing near perfect reconstruction of the original spectrum at low frequencies. The largest differences occur around 500 Hz, which might be specific for the dataset used for evaluation. The way accuracy depends on frequency can be impacted by frequency mapping. Linear mapping treats all frequencies as equally important in the least-squares fitting, while it may be beneficial to employ a frequency scale more aligned with how the human hearing system works, e.g. a logarithmic or mel scale. On the other hand, it is known that spectral localization cues are dependent mostly on the shape of pinna, which impacts only higher frequencies \cite{Algazi2001,Langendijk2002}. Alternatively, the importance of accuracy in certain frequency bands can be manipulated by the frequency weighing if needed. 

To further put the magnitude of errors in perspective, the approximation errors can be put against the measurement errors. Some research has been presented on the impact of different variables on the spectral variations between HRTF measurements \cite{Andreopoulou2015a,Andreopoulou2013}; HSH approximation error is on par or lower than the human-related variations in repeated individual HRTF measurements or the differences between HRTF sets of the same artificial head but obtained in different laboratories. What is more, the difference between SH and HSH approximations is lower than the differences between the measurements performed in the same laboratory in the span of a week (0.12 dB vs 0.17 dB). The frequency smoothing introduced by HSH approximation, although changing the original values, might also be perceived as a positive effect, since some of the distortion in the raw data might be caused by noise and imperfections of the measurement setup.


Overall, application of HSHs instead of SHs to approximate magnitude of HRTFs yields almost the same accuracy while exhibiting some additional advantages. First of all, the resulting representation is fully continuous, not only in space but also in frequency. This continuity enables retrieving discrete HRTFs of any frequency resolution without any additional processing. The most prominent advantage of application of SHs to directivity functions was thus extended to another dimension. An exemplary benefit of using HRTF model of infinite frequency resolution is the control of balance between precision and latency of binaural rendering by reading discrete HRTFs of any desired resolution.

What is more, even less amount of data is required to store a HRTF set in the HSH domain; in the analyzed example, the HSH approximation was described by 3081 coefficients, while the SH one required 81 for each of 257 frequency bins (255 excluding the first two bins that were ignored in the HSH approximation either way). The total number of SH coefficients was thus 20817 (20655), which is almost seven times more than in the HSH approximation, while retaining similar amount of perceptually relevant information. Even if SH approximation was determined at only 40 frequency bands to match the reduced frequency resolution of HSHs for $n_{\text{max}}=80$, the total number of SH coefficients would be 3240 which is slightly more than the number of HSH coefficients (3081). It is also worth noting that 40-bin HRTFs would be computationally inefficient and would need to be resampled while no such operation is needed in the HSH representation, where spectra of any resolution can be read without any extra processing. Furthermore, comparing the number of HSH coefficients (3081) to the number of raw data samples (182470), the reduction of data size is almost 60-fold with SD below 2.5 dB, while the best configuration of the HRTF compression method proposed by Shekarchi et al. reached compression ratio of only 40 for the same accuracy (to get the compression ratio of 60, SD increases to about 2.8 dB) \cite{Shekarchi2013}. Furthermore, their method is discrete in both space and frequency while the HSH representation is fully continuous. HSHs can be thus useful in designing a compression format that would include all the variability of HRTFs within a single set of coefficients, allowing them to be expressed holistically and compactly.

The downside of the HSH representation is that it requires more computational resources to determine the HSH coefficients via least-squares fitting. However, with the continuous development of processing units, this issue becomes less and less relevant.

The mathematical structure of the HSH representation allows it to be utilized not only for HRTFs but for any type of directivity function, e.g. directivity of electroacoustic devices, be they sound sources or receivers. These types of directivity functions are easier to measure and so applying HSHs as approximation tools might be of lesser value, but they can still be useful for such objects in any machine learning problems requiring proper directivity parametrization. HSHs, capturing holistic nature of the far-field directivity, seem to be better fitted for such tasks than SHs.

One of the interesting questions regarding presented method is the impact of the choice of limiting parameters $n_{\text{max}}$, $l_{\text{max}}$ and $m_{\text{max}}$ on the approximation accuracy. Increasing these limits will improve the accuracy, but the quality and quantity of the improvement remains to be determined. This is intrinsically linked with the question of what is the minimum required accuracy from the perceptual standpoint and what is the best way to achieve it. However, the presented subject is very broad and requires a thorough analysis to provide reliable answers. For this reason, it is decided to be out of scope of the present paper, but will be considered for future research.

\section{Conclusions}
\label{s:conclusions}

Within this paper, the theory of HCS, real HSHs and their application to express frequency-dependent directivity data such as HRTFs was presented. The entire computational process was described, starting from discrete HRIRs to determining HSH coefficients. Special focus was put on efficient mapping of frequency scale to an angle, which was required to express the directivity data in HCS. The location of the hyperspherical poles was leveraged to match the omnidirectionality of directivity functions such as HRTFs for low frequencies, making the HSH basis surprisingly well-suited to express such functions, given no physical motivation. 

Exemplary HSH representation was determined for KEMAR HRTFs using perceptually-motivated number of HSHs. The HSH approximation yields similar levels of accuracy to the corresponding SH one, while providing continuity in frequency and a significant reduction of required amount of data. Thanks to the continuity over both space and frequency, the HSH representation captures the holistic nature of far-field HRTF characteristics. Thus, the HSH representation can be considered an upgrade over the currently popular SH-based approach in practical applications. However, there are still many aspects to investigate, such as e.g. impact of the limiting parameters on the accuracy, exploring different frequency mapping or weighing, perceptual tests, etc.

The method described within this paper not only is a value by itself, but also presents wider possibilities of perceiving directivity functions by modeling frequency as an extra dimension, introducing coupling between their spectral and spatial properties. The HSH model can serve as a base for development of similar representations applying different 4D functions, e.g. created by merging basis functions of lower dimensions. Furthermore, this paper focuses only on the magnitude of far-field HRTFs, which can be in future complemented by the distance and phase dependences if needed.

\section*{Acknowledgements}
This research was supported by the National Science Centre, project No. 2020/37/N/ST2/00122. The author would like to thank Anna Snakowska for advice and helpful discussions as well as anonymous reviewers for their insightful suggestions.

\bibliographystyle{ieeetr}
\bibliography{biblioAA}

\end{document}